\documentclass[12pt,twocolumn]{aastex63}
\usepackage{bm}
\usepackage{ulem}

\newcommand\K{{\rm\,K}}

\newcommand{\tkedt}[2]{\textcolor{red}{{#1} {\sout{#2}}}}

\shortauthors{Kawashima et al.}
\shorttitle{}

\received{September 18, 2020}
\submitjournal{ApJ}

\begin{document}
%
\title{A Jet-Bases Emission Model of the EHT 2017 Image of M87*}

 \correspondingauthor{Tomohisa Kawashima}
 \email{kawshm@icrr.u-tokyo.ac.jp}

 \author[0000-0001-8527-0496]{Tomohisa Kawashima}
 \affil{National Astronomical Observatory of Japan, 
 2-21-1 Osawa, Mitaka, Tokyo, 181-8588, Japan}
\affil{Institute for Cosmic Ray Research, The University of Tokyo, 5-1-5 Kashiwanoha, Kashiwa, Chiba 277-8582, Japan}
 
 \author[0000-0002-7114-6010]{Kenji Toma}
 \affil{Frontier Research Institute for Interdisciplinary Sciences, Tohoku University, Sendai, 980-8578, Japan}
 \affil{Astronomical Institute, Graduate School of Science, Tohoku University, Sendai, 980-8578, Japan}

 \author[0000-0002-2709-7338]{Motoki Kino}
 \affil{National Astronomical Observatory of Japan, 
 2-21-1 Osawa, Mitaka, Tokyo, 181-8588, Japan}
 \affil{Kogakuin University of Technology \& Engineering,, Academic Support Center, 
 2665-1 Nakano, Hachioji, Tokyo 192-0015, Japan}
 
\author[0000-0002-9475-4254]{Kazunori Akiyama}
\affil{National Radio Astronomy Observatory, 520 Edgemont Road, Charlottesville, VA 22903, USA}
\affil{MIT Haystack Observatory, 99 Millstone Road, Westford, MA 01886, USA}
\affil{National Astronomical Observatory of Japan, 
2-21-1 Osawa, Mitaka, Tokyo, 181-8588, Japan}
\affil{Black Hole Initiative, Harvard University, 20 Garden Street, Cambridge, MA 02138, USA}

 \author[0000-0001-6081-2420]{Masanori Nakamura}
 \affil{National Institute of Technology, Hachinohe College, 16-1 Uwanotai, Tamonoki, Hachinohe City, Aomori 039-1192, Japan}
 \affil{Institute of Astronomy \& Astrophysics, Academia Sinica, 11F of Astronomy-Mathematics Building,
AS/NTU No. 1, Taipei 10617, Taiwan}

\author[0000-0003-1364-3761]{Kotaro Moriyama}
\affil{MIT Haystack Observatory, 99 Millstone Road, Westford, MA 01886, USA}




\begin{abstract}
We carry out general relativistic ray-tracing radiative-transfer calculations to study whether a localized emission from 
plasma rings created at the stagnation surface in the jet funnel, to which we refer as stagnation rings, can explain the ring image of M87* observed by Event Horizon Telescope (EHT) 2017.
The resultant images consist of the direct image of the stagnation rings and the ring images formed via the strong deflection by the black-hole (BH) gravity, to which we refer as "quasi photon-ring".
For the model with the BH spin $a_* = 0.99$, the direct image of the counter-jet and quasi photon-ring are almost coincident to the photon ring with diameter $\sim 40 \mu{\rm as}$, while the approaching jet shows the small ring-image inside them.
The synthetic observation image assuming the EHT 2017 array is consistent with that observed in M87*, because the array is a bit sparse to detect the inner ring image. 
This indicates that the ring image in M87* may contain the important feature of the jet bases in addition to the photon ring.
We find that forthcoming EHT observations can resolve the stagnation-ring image and may enable us to explore the plasma-injection mechanism into the jet funnel. 
\end{abstract}


\section{Introduction}

Many general relativistic (GR) magnetohydrodynamic (MHD) simulations of magnetized accretion flows onto Kerr black holes (BHs) \citep[][and references therein]{2004ApJ...611..977M,2019ApJS..243...26P} show that a BH-driven relativistic jet is realized in the magnetically-dominated funnel region via Blandford-Znajek process \citep{1977MNRAS.179..433B,2009mfca.book.....B,2016PTEP.2016f3E01T}, while the matter-dominated turbulence outside the funnel does not drive relativistic outflow \citep{2013MNRAS.436.3856S,2018ApJ...868..146N}.
In the funnel region, particle outflows are thought to originate from the stagnation surface, below which the particles fall by the BH gravity \citep{1990ApJ...363..206T,2015ApJ...801...56P}. This picture of the funnel is based on MHD, whereas the MHD condition might be broken in some parts at/below the stagnation surface \citep[e.g.,][]{2015ApJ...809...97B,2016ApJ...818...50H,
2017PhRvD..96l3006L,
2020ApJ...892...37P,2020ApJ...894...45H, 2020ApJ...902...80K}. 
Such inner regions of jets have not been probed by any observation.

The giant elliptical galaxy Messier 87 (M87) is one of the nearest radio galaxies with a prominent relativistic jet extending to several kilo-parsec scales \citep{1989ApJ...340..698O,1996ApJ...473..254S}, 
which has been studied in detail with
Very Long Baseline Interferometry (VLBI) radio observations \citep{2011Natur.477..185H,2012ApJ...745L..28A,2016A&A...595A..54M,2018ApJ...855..128W} and their theoretical modelings \citep{2013ApJ...775..118N,2014ApJ...786....5K,2015ApJ...803...30K,2018ApJ...868..146N}. 
The limb-brightening feature with superluminal blob motions at 15-86 GHz \citep{2007ApJ...668L..27K,2016ApJ...817..131H,2018ApJ...855..128W} 
which is seen down to $\sim 50\;r_g$ \citep[$r_g \equiv GM/c^2$;][]{2018A&A...616A.188K} may have hints for driving and emission mechanisms of the jet \citep{2018ApJ...868...82T,2018ApJ...868..146N,2019ApJ...877...19O}.
More inner region can be investigated with increasing frequency as the jet becomes increasingly more transparent. 
One of key questions is
whether the particle flow starts with bright emission at the stagnation surface near the jet edge 
\citep{2015ApJ...809...97B,2017ApJ...841...61A,2017ApJ...845..160P}.

The Event Horizon Telescope (EHT) observed the center of M87 in 2017 with $\sim 25\;\mu{\rm as}$ angular resolution at 230 GHz, and detected the BH shadow surrounded by bright ring-like emission \citep[][hereafter EHTC2019a, b, c, d, e, f]{2019ApJ...875L...1E,2019ApJ...875L...2E,2019ApJ...875L...3E,2019ApJ...875L...4E,2019ApJ...875L...5E,2019ApJ...875L...6E}.
In EHTC2019e, comparing the observational data to theoretical models which combine GRMHD simulations and GR ray-tracing radiative-transfer (GRRT) calculations revealed that the observed ring originates from diffuse, optically-thin synchrotron emission from thermal electrons in the accretion disk and/or “funnel wall” (i.e., the region just outside the funnel) at $r \lesssim 4 \;r_{\rm g}$. 
However, those calculations in EHTC2019e assumed no emission 
from  the  funnel region (i.e., the ratio of the magnetic to rest-mass density $\sigma > 1$), and thus its contribution to the observed ring image has not been thoroughly studied yet.
It is not clear  whether  the  emission observed at 230 GHz is dominated by the accretion flow or the jet
\citep[see also, e.g.,][]{2012MNRAS.421.1517D}.

In this \tkedt{paper}{Letter}, we build a simple model of emission in the funnel, specifically at the bottom of the stagnation surface (with no emission from the accretion disk or funnel wall), and examine whether such emission can reproduce the observed ring-like emission structure by calculating GRRT and subsequent image reconstruction assuming EHT arrays.

\section{Setup of Stagnation Ring Model and GRRT Calculations}\label{sect:model_setup}


\begin{deluxetable}{ccccc}
\tablecaption{Parameters of stagnation ring model.}\label{tab:model_parameter}
\tablewidth{10pc}
\tablehead{
\colhead{$a_*$} &  \colhead{$r_{\rm f} ~[r_{\rm g}]$} & \colhead{$n_{\rm e (nth)} ~[{\rm cm}^{-3}]$}   & 
\colhead{$\Omega ~ [r_{\rm g}/c]$} 
 & 
\colhead{$\Omega/\Omega_{\rm H}$
}}
\startdata
0.5 & 13 &  
$8.9 \times 10^{2}$
&  0.06 & $\simeq$ 0.45 \\
0.7 & 10 &  
$1.2 {\times} 10^{3}$
&  0.08 & $\simeq$ 0.36\\
0.9 &  6.5 & 
$3.5 {\times} 10^3$
&  0.15 & $\simeq$ 0.48 \\
0.99 & 4  & 
$6 {\times} 10^{3}$
&  0.2 & $\simeq$ 0.46\\
\enddata
\tablecomments{The parameters $B = 50$ [G], $r_{\rm size} = 0.5$ $[r_{\rm g}]$, $\gamma_{\rm min} = 50$, $\gamma_{\rm max} = 5\times10^3$, and $p = 3.5$ are used in the all models in this work.
}
\end{deluxetable}

\begin{figure}[!t]
\centering
\includegraphics[width=1.0\columnwidth]{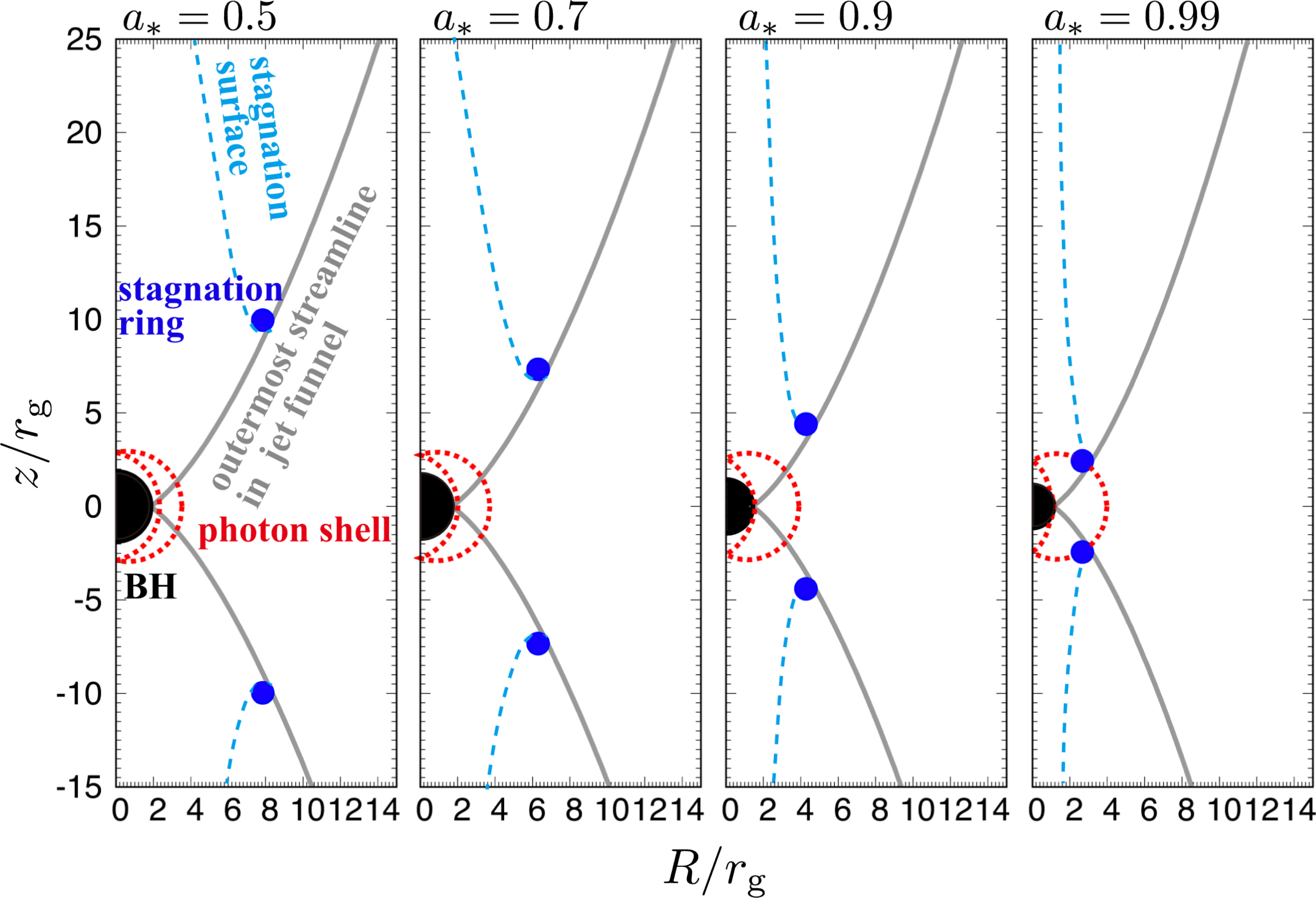}
\caption{Location and geometry of the stagnation rings.
The filled blue circles display the stagnation rings. 
The gray curves present  outermost streamline in the jet funnel described by Equation (\ref{eq:stream}).
The cyan dashed lines describe the rough position of the stagnation surface
\citep[more precise and detailed structure is shown in Figure 14 in][]{2018ApJ...868..146N}.
 The red dotted-lines display the innermost and outermost photon spheres \citep[][, EHTC 2019e]{2003GReGr..35.1909T}, to which is referred as the photon shell \citep{2020SciA....6.1310J}. 
 The photon spheres for the observer with  $i=163^{\circ}$ exist inside the photon shell.}
\label{fig:position_stag_ring}
\end{figure}

We compute the images of the stagnation ring with non-thermal electrons around the Kerr BH, by using a
GRRT 
code \texttt{RAIKOU} \citep[][Kawashima et al. in prep.]{2019ApJ...878...27K}.
We set the BH mass $M = 6.5 {\times} 10^{9}M_{\odot}$ (EHTC2019f) and the BH-spin parameter $a_* = 0.5, 0.7, 0.9$, and 0.99.
The observer screen 
with the field of view $160\mu{\rm as} \times 160\mu{\rm as}$  divided by $2400 {\times} 2400$ pixels
is located at $10^4 r_{\rm g}$ with viewing angle $i=163^{\circ}$.
We assume the distance of M87* to be $D=16.7$ Mpc \citep{2010A&A...524A..71B}.

\begin{figure*}[!t]
\centering
\includegraphics[width=0.99\textwidth]{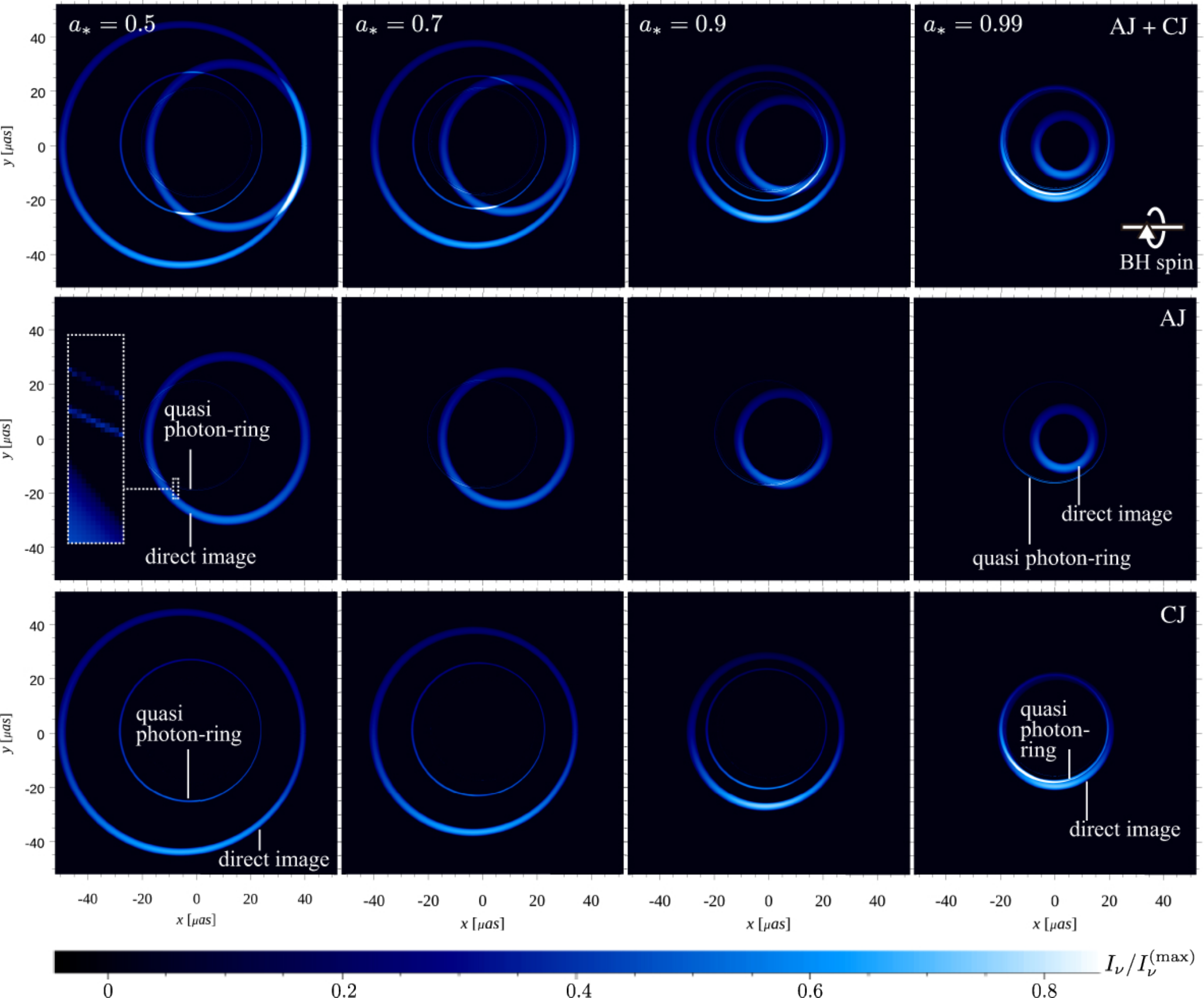}
\caption{Stagnation-ring images  with both of the approaching-jet (AJ) and counter-jet (CJ) emission (top), AJ only (middle), and CJ only (bottom) with PA$=270^{\circ}$. 
The color represents the intensity normalized by the maximum intensity including both of the approaching- and counter-jet for each BH-spin parameter.
Since the quasi photon-ring is too thin and dim in the palel of AJ emission with $a_*=0.5$, we insert the rectangular box showing the magnified view of a part of the quasi photon ring and the direct image of the approaching jet, in which the sub-structure of the quasi-photon ring can be also found.
}
\label{fig:image-non-thermal}
\end{figure*}

First of all, we present the location and the structure of the stagnation ring (see Fig. \ref{fig:position_stag_ring}).
We set 
plasmas at the bottoms of the stagnation surfaces in the funnel. 
This is because the breakdown of MHD condition could occur there \citep{2015ApJ...809...97B}.
The breakdown of MHD condition is the emergence of electric field parallel to the magnetic field, which is caused  by  the  strong  magnetization  inside  the jet funnel suppressing diffusion of charged particles from  the  accretion  flow  and the  resultant  low  density  plasma insufficient to screen the electric field.
More particles are expected to be created via, e.g., inverse-Compton pair-catastrophe, at regions with stronger magnetic fields which are closer to the BH.
The position of the stagnation-ring center locates inside 
the outermost streamline in the jet funnel, which can be represented by the magnetic stream function \citep{2008MNRAS.388..551T} being connected with outer horizon of the BHs on the equatorial plane: 
\begin{equation}
    \Psi(r_{\rm f}, \theta_{\rm f}) = 
    \left(
    \frac{r_{\rm f}}
    {r_{\rm H}}
    \right)^{\kappa}(1- \cos{\theta}_{\rm f}) =
     1,
\label{eq:stream}     
\end{equation}
where $(r_{\rm f}, \theta_{\rm f})$ describes the outermost streamline in the jet funnel and $r_{\rm H} = r_{\rm g}(1 +  \sqrt{1-a_{*}^{2}})$
is the outer horizon radius of the Kerr BHs.
We set $\kappa = 0.75$ in such a way that the magnetic-streamline shape is consistent with the VLBI observations \citep{2016ApJ...817..131H, 2018A&A...616A.188K}, and choose $r_{\rm f}/r_{\rm g}$ as summarized in Table \ref{tab:model_parameter} to be consistent with GRMHD simulations \citep{2018ApJ...868..146N}
\footnote{In \cite{2018ApJ...868..146N}, the accretion flow is in the semi-MAD state, which is an intermediate state between SANE (Standard And Normal Evolution)  and MAD (Magnetically Arrested Disk), see  \cite{2012MNRAS.426.3241N, 2013MNRAS.436.3856S, 2015ASSL..414...45T}; EHTC2019e, and references therein for the detail of SANE and MAD.
The SANE and MAD are weakly and strongly magnetized states, which are defined by the dimensionless magnetic flux threading the event horizon $\phi_{\rm BH} ~{\sim} 1 $ and $\sim 15$, respectively.
Here, $\phi_{\rm BH} = \Phi_{\rm BH} / \sqrt{{\dot M}_{\rm BH} r_{\rm g} c}$, and $\Phi_{\rm BH} = (1/2) \int_\theta \int_{\varphi} |B^r| dA_{\theta \varphi}$, $dA_{\theta \varphi}$ is an area element in the $\theta$--$\varphi$ plane, and ${\dot M}_{\rm BH}$ is the mass accretion rate onto the BH.
The magnetic field in MAD is so strong that it  obstructs the steady infall of plasma and results in the accretion with strong time-variability
\citep{2003ApJ...592.1042I, 2003PASJ...55L..69N}, see also \cite{1974Ap&SS..28...45B, 1976Ap&SS..42..401B}.
The strong magnetic field in MAD leads to  formation of the wider jet funnel and also the higher jet efficiency (i.e., higher ratio of jet-power to accretion-power) due to more efficient Blandford-Znajek process than that in SANE.
}



The radius of cross-section circle of the stagnation rings (the filled blue circles in Fig. \ref{fig:position_stag_ring}) is set to be $r_{\rm size} = 0.5 r_{\rm g}$, being in rough agreement with  the estimated scale length of the emission region inside the M87 jet. 
The synchrotron cooling timescale $t_{\rm syn}$ limits the scale 
length of the emission region as $\ell_{\rm syn} = c t_{\rm syn}$
%
$
= 3m_{e}c^{2}/4\sigma_{\rm T}U_{B}\gamma_{\rm e} 
~ \sim ~ 0.3 r_{\rm g}
({B}/50~{\rm G})^{-3/2}
(\nu_{\rm syn}/230~{\rm GHz})^{-1/2},
$
where we have assumed that the Lorentz factor of bulk motion is $\sim 1$ at the stagnation surface, and 
$\nu_{\rm syn} = eB\gamma_{\rm e}^2/2\pi m_e c$.

Physical quantities of stagnation rings are as follows.
We set the magnetic-field strength
 $B = 50\;$ G \citep{2015ApJ...803...30K}. 
The energy spectrum of the non-thermal electrons are assumed to be $\propto {\gamma}^{-p}$ in the range $50 \le \gamma \le 5 \times 10^3$, where $\gamma$ is the Lorentz factor of the electrons and the power-law index is $p = 3.5$.  
Here, the minimum Lorentz factor is chosen to be $\nu_{\rm syn} \sim 230$ GHz, being consistent with the parameter range in \citet{2012MNRAS.421.1517D}. The maximum Lorentz factor is set to be so high that it does not affect the results at 230 GHz.
We set the number density of the non-thermal electrons 
as shown in Table \ref{tab:model_parameter},
in such a way that the resultant radiative flux at 230 GHz to be ${\simeq} ~0.6 $Jy
(EHTC 2019d).
Normalized angular velocity of the stagnation ring is set as described in Table \ref{tab:model_parameter}.
Here, the angular velocity $\Omega = u^{\varphi}/u^{t}$ is evaluated by using the azimuthal and time component of the four-velocity $u^{\varphi}$ and $u^t$ measured in the observer frame at the stagnation surface in GRMHD simulations \citep{2018ApJ...868..146N},
which 
 will be almost equivalent with the angular velocity of the BH magnetosphere, since the radial (and poloidal) velocity is zero at the stagnation surface.
We note that $\Omega/\Omega_{\rm H} ~{\simeq}$
0.5 is also roughly consistent with those in another type of GRMHD simulations solving inside the jet funnel \citep{2010ApJ...711...50T}, where $\Omega_{\rm H} = a_* c/ 2r_{\rm H}$.

We calculate the GRRT images of the stagnation-ring model at 230 GHz.
We assume that it is vacuum outside the stagnation ring 
to focus on the possibility that stagnation ring mimic the ring-like image observed in M87 without the uncertainty of the accretion flow emission.
The synchrotron emission and absorption via the non-thermal electrons are incorporated
 as described in \cite{2016MNRAS.462..115D},
in which the coefficient of emissivity and the absorption are numerated and tabulated without assuming $\gamma_{\rm min}^2 {\nu}_{\rm p} \ll {\nu} \ll \gamma_{\rm max}^2 {\nu}_{\rm p}$, 
 where $\nu_{\rm p} = 3 e B \sin \theta_{\rm B} /4 \pi m_{\rm e} c $.
Following some works based on semi-analitic models \citep[e.g.,][]{2016ApJ...831....4P}, we fix the angle between the ray and the magnetic field to be $\pi/6$ for 
simplicity.
\section{Simulated Stagnation Ring Images and Comparison with the Ring Images of M87* }\label{sect:result}


The resulting image of the stagnation ring is shown in Figure \ref{fig:image-non-thermal}.
The top panels present the total ring-images including both of the approaching- and counter-jet.
The position angle of the jet is assumed to be $270^{\circ}$, i.e.,  the observer is in the the West (right) direction in the screen.
It is shown that the diameter of the all ring-images decreases with increase of the BH spin, as a consequence of the appearance of the stagnation ring closer to the BH when the BH spin is higher.
Importantly, for $a_* = 0.99$, the stagnation-ring image in the counter-jet region (i.e., the outer ring) almost coincide with the photon ring with diameter ${\sim} 40 {\mu} {\rm as}$, which is consistent with the observed ring diameter in M87*.
We also note that that the small ring via the approaching jet emission appears inside the ${\sim} 40 \mu {\rm as}$ ring.

In order to understand these complicated ring feature,
 we decomposed the images into those from the approaching- and counter-jet emission, as shown in the middle and bottom panels in Figure \ref{fig:image-non-thermal}, respectively \citep[see also Appendix A in][for more simplified model with $a_* = 0$]{2019A&A...632A...2D}.

\begin{figure*}[t]
\centering
\includegraphics[width=0.99\textwidth]{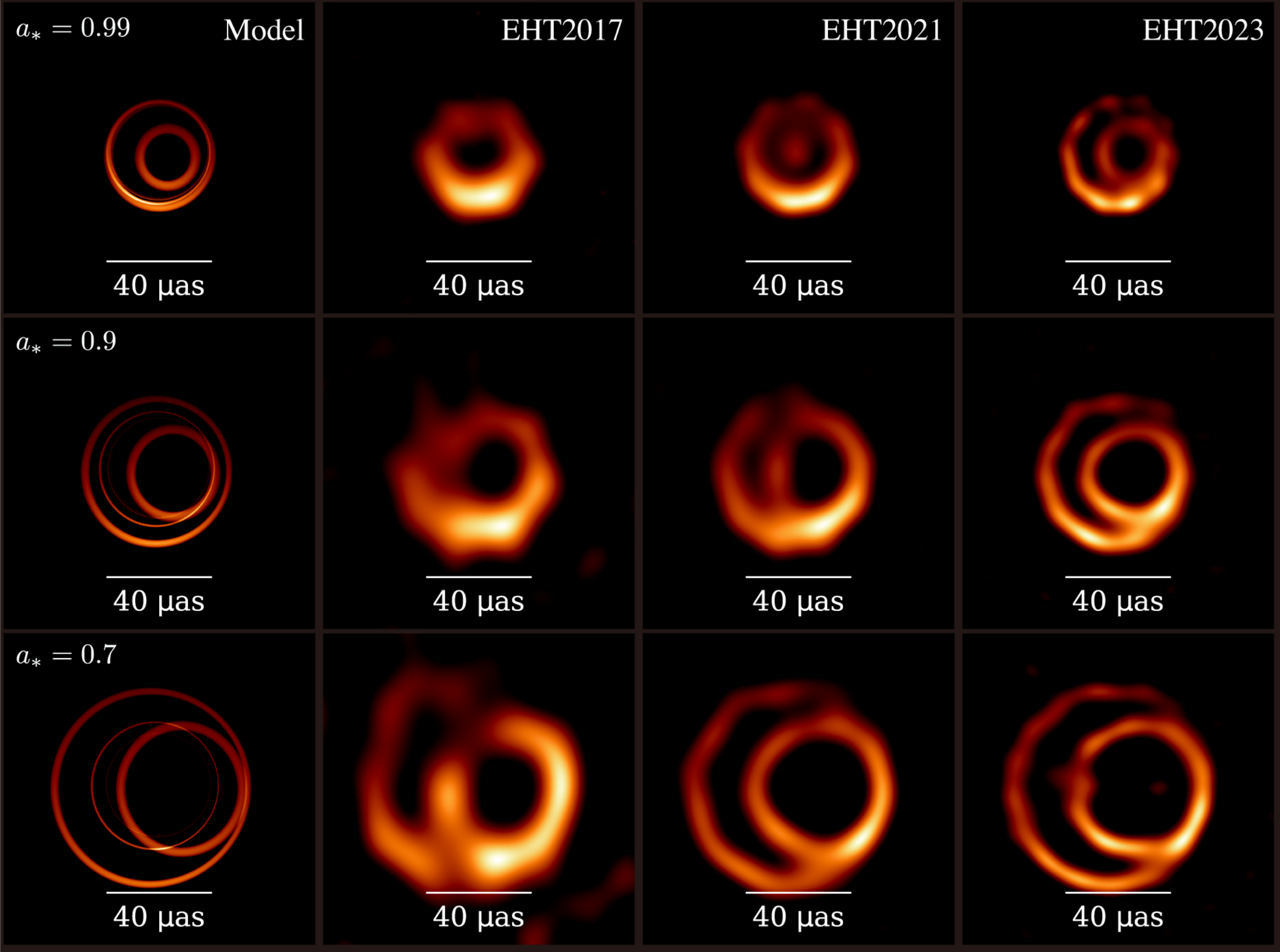}
\caption{Reconstructed image of models with $a_* = 0.99$(top), 0.9 (middle), and 0.7 (bottom) using \texttt{SMILI}. 
The first column displays the theoretical model as a reference. The second, third, and fourth columns present the reconstructed images assuming the array of EHT 2017, 2021, and 2023, respectively.}
\label{fig:image-reconstruct}
\end{figure*}

The bottom panels in Figure \ref{fig:image-non-thermal} display the images of photons emitted from the counter-jet region.
The separated ring images (i.e., the outer-broad and the inner-narrow rings) appear in all the models except the model with $a_* = 0.99$.
The outer-broad ring is the (gravitationally-lensed) direct emission images of the stagnation ring. 
The inner rings are formed by the photons which turned around the BH after the emission from the stagnation ring. 
The images are nearly identical to the photon ring, however they are slightly larger than that.
This is because the photons propagate in the region slightly outside the photon spheres 
after their localized emission from the stagnation ring\footnote{Of course, the inner rings are composed of multiple sub-ring images, which is formed by photons turn around the BH less and more than one orbit. When the photons rotate more, the image becomes more similar to the photon ring  with lower resulting radiative flux \citep[see, e.g.,][]{1979A&A....75..228L, 2020SciA....6.1310J}. 
In this study, we focus on the most luminous one formed by the photons turning around the BH less than one orbit, so that the diameter of the ring-image is slightly larger than the photon ring, especially for the lower BH-spin models.}
We refer to this ring image as a "quasi photon-ring".
The quasi photon-rings asymptotically become coincident with the photon ring as the BH spin increases, since the stagnation ring locates closer to the BH.
The size of the gap between quasi photon-rings and the direct ring-images decreases as the BH spin increases, since the diameter of the quasi photon-ring more weakly depends on the BH spin than the outer direct images.
Finally, for $a_* = 0.99$, these two rings  merge and are almost coincident with the photon ring, because a part of the stagnation rings overlaps to the photon shell (Fig. \ref{fig:position_stag_ring}).

In the middle panels, the approaching-jet emission also shows the direct emission image of the stagnation ring  and the quasi photon-rings.
The direct emission image appears in the West direction in the screen (i.e., close to the observer).
As is the case with the counter-jet,  the diameter of the direct ring-images decreases more drastically than the quasi photon-rings with increasing the BH spin.
For the model with $a_* = 0.99$, the direct ring-image of the approaching jet appears inside the (quasi) photon-rings.

As a consequence of the relativistic Doppler effect due to the rotation of the stagnation ring with the magnetosphere-rotation velocity, the rings become bright on the South side.
This effect becomes more significant as the BH spin increases, because the magnetosphere rotates faster.

Next, we show the results of synthetic observation of our theoretical images at 230\,GHz, assuming EHT arrays from the past (EHT 2017) to the future ones in Fig. \ref{fig:image-reconstruct}.
The synthetic observational data are created with the \texttt{eht-imaging} library \citep{2016ApJ...829...11C, 2018ApJ...857...23C} and imaged with \texttt{SMILI} \citep{2017AJ....153..159A, 2017ApJ...838....1A}.
We here considered three array configurations: the EHT2017 array with seven stations at five geographic sites, the EHT2021 array with the additional three stations at Kitt Peak, Plateau de Bure and Greenland 
(see EHTC2019b for details) and the EHT2023 array with the addition of 345\,GHz coverages qualitatively simulating the improvement provided by multi-frequency synthesis. We adopted the nominal sensitivities and atmospheric conditions of telescopes at 230\,GHz (EHTC2019b). Images were reconstructed with $\ell_1$+TSV regularizations 
(e.g., EHTC2019d)
providing reasonable fits ($\chi _\nu ^2 \sim 1.0$) for all three models.

For the model with $a_* = 0.99$, the synthetic observation image with the EHT2017 array quantitatively agrees with the images observed in M87*: the diameter of the ring is $\sim 40\mu$as and the brightness asymmetry in the ring appears (i.e., the South part is roughly 2 times brighter than than the North one).
The inner ring (i.e., approaching jet image) is not observed because the EHT2017 array has still sparse configuration.
The outer ring consists of the photon ring and the stagnation-ring image in the counter-jet region overlapping the photon ring, i.e., its diameter is ${\sim} 40\mu$as.
These are the reason why the resultant synthetic image coincides with the observed ring image in M87*.

On the other hand, the synthetic images of the models with the other spin parameters ($a_* =0.7$ and 0.9) are not similar to the observed image. The theoretical images of these parameters show the ring of counter-jet with diameter significantly larger than that of photon ring, and these large ring images are well reconstructed in the synthetic images. For the model with $a_* =0.7$, the inner ring is also reconstructed, i.e., the double-ring structure appears in the synthetic images.
These features are not found in the M87* ring-images (EHT2019c,d,f), so that the models with $a_{*} \le 0.9$ are disfavored.

As is shown in the third and fourth columns in Fig. \ref{fig:image-reconstruct}, future EHT observations can identify the existence of the stagnation ring. Here, we focus on the model with $a_* = 0.99$. With the EHT2021 array, a faint feature of the inner ring can be detected.
However, it a faint spot image and more clear images will be required to certificate the appearance of the stagnation ring.
If we assume the EHT2023 array, the resolution of the image is drastically improved  thanks to 345\,GHz coverages and one can successfully identify the inner ring (i.e., approaching jet) in the image.
This indicates that forthcoming EHT observations will enable us to test the models, and furthermore, to explore the plasma-injection mechanism of the relativistic jets.

\section{Summary and Discussion}\label{sect:summary}

We carried out GRRT calculations to study whether the localized emission from stagnation ring, which is the 
plasma ring created at the stagnation surface in the jet funnel, can reproduce the ring-like image of M87*.
We found that the resulting images consist of the direct image of the stagnation ring and the images formed via the strong deflection of the ray, to which we refer as quasi photon-ring.
The diameter of the ring-images drastically decrease except the quasi photon-ring with increasing the BH spin, because the stagnation surface appears in the region closer to the BH when the BH spin is higher. 

For the model with $a_* = 0.99$, direct ring-image by stagnation ring in the counter-jet region and the quasi photon-ring are almost coincident with the photon ring.
The inner ring, which is the direct image of the stagnation ring in the approaching jet, appears inside these $\sim 40 \mu {\rm as}$ rings.
Importantly, the inner ring is difficult to be resolved by using EHT 2017 array.
This indicates that the asymmetric ring image observed in M87* may include a direct-emission image from the jet basis in addition to the photon-ring image.
Forthcoming EHT observations of M87*
can resolve the inner ring feature. If the inner-ring image would be detected, it may enable us to study the plasma-injection and launching mechanism of the relativistic jet.\footnote{The future EHT observation may also distinct the model of electron distribution-function \citep{2020A&A...636A...5R} and non-Kerr objects, e.g., wormholes \citep{2020PhRvD.102h4044W} and boson stars \citep{2020arXiv200209226V}.}


One may think that the approaching-jet emission would reproduce the ring image with $\sim 40 \mu{\rm as}$ for the models with the BH spin between $a_* = 0.7$ and 0.9 (see Fig. \ref{fig:image-non-thermal}), if the counter-jet was obscured by the accretion flow. 
However, this situation will be difficult to be realized.
This is because the remarkable absorption by the accretion flow will also result in the significant emission from itself.

The image morphology of our $a_* = 0.99$ model is similar to that of a SANE model  with $a_*=0.94$ (Fig. 2 in EHTC2019e). 
The resembling images are a consequence of the similar location the emission region, i.e., that inside the mildly wide funnel of semi-MAD (our model) and outside the narrow funnel of SANE (EHTC2019e).
Importantly, their SANE models passed the reduced $\chi^2$ test for imaging, which means that our reproduction of the image-features of M87* in the reconstructed image 
is reasonable.
It should be mentioned that our stagnation ring model is rather motivated to be applied to a case of the emission from the highly magnetized funnel in MAD (or semi-MAD) models, while no emission is assumed in EHT2019e. 
Simultaneous calculations of images of stagnation ring and the accretion disks in MAD or semi-MAD state remain as future works. The calculations may result in an additional faint, blurred disk-image and the time-variation of the bright region due to the magnetic reconnenction and/or magnetic interchange modes. 

The ratio of energy density of 
electrons
to magnetic field is 
$U_{\rm e}/U_B = 6.1{\times}10^{-4}, 
8.2{\times}10^{-4}, 
2.4{\times}10^{-3}$, and $4.1{\times}10^{-3}$
for $a_* = 0.5$, 0.7, 0.9, and 0.99, respectively,
where $U_{\rm e} = \int_{\gamma_{\rm e, min}}^{\gamma_{\rm e, max}} d\gamma_{\rm e} [\gamma_{\rm e} m_{\rm e}c^2 n_{\rm e, nth}(1-p) \gamma_{\rm e}^{-p}/(\gamma_{\rm e, max}^{1-p} - \gamma_{\rm e, min}^{1-p})]$.
This is consistent with the magnetic-energy-dominant M87-jet picture \citep{2015ApJ...803...30K}.
On the other hand, \cite{2020MNRAS.492.5354M} demonstrated that particle energy dominates at the emission site of the very high energy (VHE) $\gamma$-ray in M87.
Unified picture of the jet which simultaneously explains the radio and VHE $\gamma$-ray emission may be addressed by considering multi-zone disk-jet models.

In this paper, we simply set the number density 
of 
electrons
to reproduce the observed flux in M87* at 230 GHz and assumed that the bottom of the stagnation surface is bright and the other part of the surface is dim.
It should be noted that the number density of $e^{\pm}$ plasma 
is higher than the Goldreich-Julian density 
$(\sim ~ \Omega B/2\pi e c$  $\sim$  $a_{*}B/8\pi e r_{\rm H}$ $\sim$ $10^{-5}{\rm cm}^{-3}$ for $a_* = 0.99$, where $e$ is the elementary charge). In order to inject the high density $e^{\pm}$, the $\gamma \gamma$ pair-production \citep{2011ApJ...735....9M},  an inverse-Compton pair-catastrophe amplified by a post-gap cascade \citep{2015ApJ...809...97B} and further processes \citep[e.g., $e^{\pm}$ injection-processes initiated by the proton acceleration in magnetically arrested disks,][]{2020arXiv200313173K} would be needed.
Alternatively, magnetic reconnection near 
the jet bases may inject electron-proton
plasma from the accretion flow into the jet funnel.
These processes may also inject the plasma into the other part of the stagnation surface. 
Exploring the injection mechanism of
plasma 
will be addressed in future work.

\acknowledgements{We thank EHT Collaboration members (especially H.-Y. Pu and C.M. Fromm, M., Wielgus, M. Janssen), K. Ioka and K. Ohsuga for the useful comments and discussion. 
The numerical simulations were carried out on the XC50 at the Center for
  Computational Astrophysics, National Astronomical Observatory of
   Japan.
This work was supported by JSPS KAKENHI grant Nos. JP18K13594, 19H01908, 19H01906 (T.K.), JP18H01245 (K.T.), JP18K03656 and JP18H03721 (M.K.).
This work was also supported by MEXT as “Program for Promoting Researches on the
Supercomputer Fugaku” (High-energy astrophysical phenomena in black holes and supernovae).
   }

\bibliography{GRRT.bib,GRRT2.bib}{}

\end{document}